\documentclass[aoas,preprint]{imsart}

\arxiv{arXiv:1607.07515}

\usepackage[latin1]{inputenc}
\usepackage{amsmath}
\usepackage{amsfonts}
\usepackage{amssymb}
\usepackage[pdftex]{graphicx}
\usepackage[colorlinks,citecolor=blue,urlcolor=blue]{hyperref}
\usepackage[margin=1in]{geometry}
\usepackage{natbib}
\usepackage{algorithm}
\usepackage{algorithmic}
\usepackage{listings}
\usepackage{dcolumn}
\usepackage{color, colortb}
\usepackage{multirow}
\usepackage{subcaption,siunitx,booktabs}
\usepackage{lscape}

\begin{document}
\begin{frontmatter}
\title{Single Stage Prediction with Embedded Topic Modeling of Online Reviews for Mobile App Management}
\runtitle{Single Stage Prediction with Mobile App Online Reviews}

\begin{aug}
\author{\fnms{Shawn} \snm{Mankad}\thanksref{m1}\ead[label=e1]{smankad@cornell.edu}},
\author{\fnms{Shengli} \snm{Hu}\thanksref{m1}\ead[label=e2]{sh2264@cornell.edu}}, 
\and \author{\fnms{Anandasivam} \snm{Gopal}\thanksref{m2}\ead[label=e3]{agopal@rhsmith.umd.edu}}

%\thankstext{m1}{}

\runauthor{Mankad, Hu, and Gopal}

\affiliation{Cornell University\thanksmark{m1} and University of Maryland\thanksmark{m2}}

\address{Shawn Mankad and Shengli Hu\\
Operations, Technology and Information Management\\
Cornell University\\
Ithaca, NY 14850\\ 
\phantom{E-mail:\ }\printead{e1}\\
\phantom{E-mail:\ }\printead{e2}}

\address{Anandasivam Gopal\\
Department of Decisions, Operations \& Information Technology \\
University of Maryland \\
College Park, MD 20910 \\ 
\phantom{E-mail:\ }\printead{e3}}
\end{aug}

\begin{abstract}
Mobile apps are one of the building blocks of the mobile digital economy. A differentiating feature of mobile 
apps to traditional enterprise software is online reviews, which are available on app
marketplaces and represent a valuable source of consumer
feedback on the app. 
We create a supervised topic modeling approach for app 
developers to use mobile reviews as useful sources of quality and
customer feedback, thereby complementing traditional software
testing. The approach is based on 
a constrained matrix
  factorization that leverages the relationship between term
  frequency and a given response variable in addition to
  co-occurrences between terms to recover topics that are both
  predictive of consumer sentiment and useful for understanding
  the underlying textual themes. The factorization is combined with ordinal regression to provide guidance 
from online reviews on a single app's performance as well as 
systematically compare different apps over time for 
benchmarking of features and consumer sentiment.
We apply our approach using a dataset of over 100,000 mobile reviews over several years for three of the most popular  
online travel agent apps from the iTunes and Google Play marketplaces.
\end{abstract}



\end{frontmatter}

\section{Introduction}
Mobile commerce is expected to reach \$250 billion by 2020 \citep{MobileBusinessInsights},
and through the increasing prevalence of smartphones, has already
started to significantly influence all forms of economic
activity. Increasingly, the mobile ecosystem is gaining significant
attention from enterprises that are porting many of their standardized
enterprise-based software functionalities to mobile platforms
\citep{serrano2013mobile}.  The rise of tablets and smartphones,
combined with the corresponding drop in PC-based traffic on the
Internet \citep{ABIResearch},
suggests that most enterprises will need to consider ``mobile'' as an
important part of their service portfolio. A central part of this
move to the mobile ecosystem is, of course, the {\em mobile app}.

Mobile apps are software products that are typically embedded in the
native operating system of the mobile device, link to various wireless
telecommunication protocols for communication, and offer specific
forms of services to the consumer
\citep{wasserman2010software, krishnan2000empirical}. One critical issue
faced by all software development teams is that of software quality
\citep{pressman2005software}, leading to the quality of experience for
the user \citep{kan1994software}. The issue of quality of experience,
based on the underlying functionality provided by the mobile app, is
of particular importance in the mobile context
\citep{ickin2012factors}, especially as service industries increase
their presence in this sphere. Poor quality of experience on the
mobile app can damage the underlying brand \citep{anthes2011invasion},
alienate rewards customers and increase defections to competitors for
more casual users, thus reducing revenues. 
These issues are also faced in enterprise software development contexts, where quality and the customer experience are particularly critical. 
To meet these requirements, mature software firms spend considerable time and effort in surveying customers and developing theoretical models of software 
quality and customer requirements before-hand \citep{parasuraman1988servqual, pressman2005software}.

In contrast to these organizational efforts to manage quality and customer requirements, however, the mobile 
developer has access to a significant quantity of feedback on the
quality of experience from the app through the channel of {\em online
  reviews}. Online reviews provide the development team with readily and easily
accessible feedback on the quality of experience from using the app,
while also influencing other potential customers' download
decisions. Moreover, useful information in such reviews are often
found in the text, rather than simply the overall rating for the
app. Thus, an arguably easy approach to understanding user-perceived quality and satisfaction with a mobile app may be to simply manually read the related online reviews and incorporate this understanding into the app development process. However, this approach poses several challenges. First, online reviews are characterized by high volume and diversity of opinions, making it harder to parse out the truly important feedback from non-diagnostic information \citep{godes2004using}. Second, they are driven by significant individual biases and idiosyncrasies, thereby making it risky to base quality improvement initiatives on single reviews or reviewers  \citep{li2008self,
  chen2013temporal, chen2014ar}. Finally, reading and absorbing all reviews associated with an app is infeasible simply due to volume, given the number of apps that are available on the marketplace, the number of reviews that are generated per app, and the rate at which new reviews are added, which is at an increasing rate \citep{lim2015investigating}.

Researchers at the intersection of software engineering 
and unstructured data analysis have developed methodologies to help the app development teams tap
into this useful source of collective information to extract specific
insights that may guide future development work on the app (see \cite{bavota2016mining} for a comprehensive survey). For example, 
\cite{chen2014ar} developed a decision support tool 
to automatically filter and 
rank informative reviews that leverages topic modeling 
techniques, sentiment, and classification algorithms. 
\cite{iacob2013retrieving, panichella2015can} and \cite{maalej2015bug} use a combination of linguistic pattern matching rules, topic modeling,  
and classification algorithms to classify reviews into different categories, like 
feature requests and problem discovery, that developers can use to filter for informative reviews. 
\cite{galvis2013analysis} applied topic modeling to app store
reviews to capture the underlying consumer sentiment at a given moment in time. 
Similarly, \cite{fu2013people} perform regularized regression with word frequencies as covariates to identify terms with 
strong sentiment that guide subsequent topic modeling of app reviews. 
The authors aggregate their findings over time to gain insight into a single app as well as all apps in the market. 

This work extends this literature to help understand the evolution of consumer sentiment over time while benchmarking apps against their competitors 
by systematically incorporating time effects and the competitive landscape into a supervised topic modeling framework that estimates 
the impact of certain discussion themes on the customer experience. 
Our data contains online reviews from the iTunes and Google Play marketplaces for three firms at the heart of the travel ecosystem 
in the United States, namely Expedia, Kayak, and TripAdvisor. All three of these firms provide apps that are free, and are aimed
at frequent travelers, with functionality for search, managing reservations, accessing promotions, 
logging into travel accounts, reviewing travel activities, and so on. 

\begin{figure}
\centerline{\includegraphics[width=0.75\columnwidth, trim=0cm 0.25cm 0cm 0cm, clip=true]{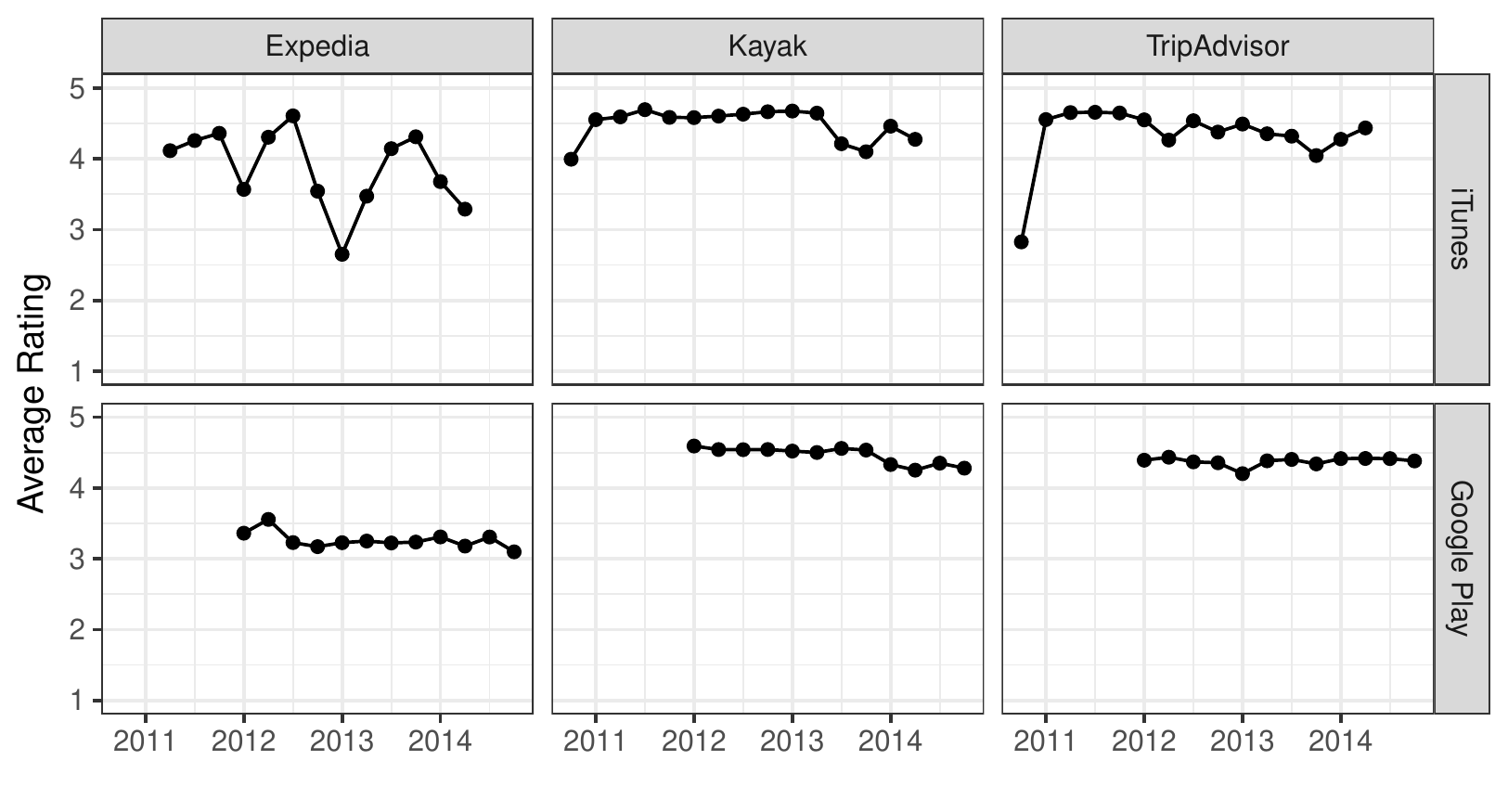}}
\caption{Average app rating over each year-quarter, by mobile app and platform.}
\label{fig:tsOverview}
\end{figure}

Figure~\ref{fig:tsOverview} shows that the time-series of 
average star ratings for each of the apps evolves over time as new versions are released. As an illustrative example, important issues for 
Expedia's managerial and development teams heading into 2013 (if not sooner) would be to understand {\em why} ratings have trended downwards 
on the iTunes platform and how consumer discussion compares to competing firms, so that appropriate remedial action can be taken to improve their positioning in the mobile marketplace. 

The main idea behind our approach is that features can be derived from
the text not only by considering the co-occurrences between terms in
reviews, but also with the observed association between term usage
and star ratings -- the response variable of interest. Thus, by using a constrained
matrix factorization embedded within an ordinal regression model, we leverage the relationship between
terms and the response variable to recover topics that are predictive
of the outcome of interest in addition to being useful for
understanding the underlying textual themes. 
The model is flexible enough to analyze multiple apps 
around common topics with evolving regression coefficients 
as new app versions are released to the public. 
These are important and novel extensions with respect to the topic modeling literature, since they allows managers and development teams 
to go beyond a static summary of the review corpus associated with an app to systematically compare different apps over time for benchmarking of features and consumer sentiment. 
%For example, we can identify using our methodology that software bugs (e.g., API errors) and user-interface are negatively %impacting the rating for Kayak in 2014. 
By pinpointing the causes of user dissatisfaction, a manager or development team can steer future development effort appropriately while ensuring a match between the user experience and the appropriate development effort by the development team.

%These functionalities need to be fixed or improved for certain apps in future releases.
%Historically to arrive at such conclusions, enterprises have relied on traditional
%software testing for quality control, which can be expensive and labor-intensive
%\citep{pressman2005software}. The mining of online reviews offers a low-cost and efficient alternative path  
%to understanding how the
%customer experience changes as new app versions are released. 

Expedia, Kayak, and TripAdvisor were three of the most reviewed travel apps at the time of collecting the data, which is 
comprised of 104,816 English reviews across a total of 162 
different versions of these apps representing the full history of these 
apps from their introduction to the iTunes and Google Play marketplaces until November 2014.  
Even in this specific context, where we limit our attention
to a particular industry and trio of apps, we see
that there are over a 1,000 reviews per app per year, with
even more reviews to be considered if the developer were 
interested in examining the reviews of competitor apps as
well, thus underscoring the need for a statistical and semi-automated approach.

%\begin{table}
%\begin{tabular}{l|r|r|r|r}
%Variable & GP-K & i-K & GP-T & i-T \\\hline
%\multicolumn{2}{l|}{GP: Google Play} &\multicolumn{2}{l}{K: Kayak} \\
%\multicolumn{2}{l|}{i: iTunes} &\multicolumn{2}{l}{T: TripAdvisor} \\ \hline
%Number of Reviews & 18873 & 12215 & 30838 & 19365 \\
%Rating & 4.488 & 4.585 & 4.393 & 4.485 \\
% & (0.980) & (0.925) & (1.084) & (0.996) \\
%Word Count & 8.887 & 12.296 & 10.967 & 20.165 \\
%& (11.218) & (15.507) & (13.466) & (23.692) \\
%\end{tabular}
%\caption{Summary Statistics (Mean, SD in paren) by App and Platform}
%\label{table:introstats}
%\end{table}

The next section presents in detail the proposed models and estimation framework followed by a review of competing methods in Section~\ref{sec:litreview}. 
Through a detailed simulation study under different generative models (Section~\ref{sec:simulation}) as well as with the iTunes and Google Play data (Section~\ref{sec:results}), 
we show that the proposed model 
performs favorably when compared to competing methods for out
of sample predictions and topic interpretability. 
We also use the results of the model to characterize 
and contrast the apps over time. The paper concludes with a
short discussion on the overall findings, the limitations of our work,
and directions for future research in
Section~\ref{sec:conclusion}.

\section{Single Stage Predictions with Matrix Factorization} \label{sec:model}

Prior work in the domain of text analytics and online
reviews \citep{cao2011exploring, galvis2013analysis, tirunillai2014mining, abrahams2015integrated,  MankadSS} has followed a two-stage approach, 
where one first derives text features through topic modeling and
subsequently applies linear regression or another statistical model
for prediction and inference. In principle there are many ways to
perform this two-stage procedure, both in terms of generating text
features and properly combining them within a statistical model. 
We address this issue by integrating both steps together using a
matrix factorization framework. The problem we focus on is prediction
and explanation of a response variable when given a set of
documents. Formally, let $X\in\mathbb{R}_{+}^{n\times p}$ be a
document term matrix with $n$ documents on the rows and $p$ terms on
the columns. Let $Y \in \mathbb{R}^{n\times 1}$ be a response
vector. Though in our application, $Y=\{1,2,3,4,5\}^{n}$ will be composed of online review scores
for apps on iTunes and Google Play, which are better modeled with an ordered multinomial distribution, we begin by solving in a novel way the case when 
the response variable is normally distributed and extend in Section~\ref{sec:multinomial} to the ordinal regression setting. 

The objective function for the proposed factorization is 
\begin{eqnarray} \label{eqn:obj}
&\min_{\Lambda, \beta}& ||Y - X\Lambda\beta||_{2}^{2} \\
&\text{ subject to }& (\Lambda)_{ij} \ge 0 \text{ for all }  i,j. \nonumber 
\end{eqnarray}
The $p \times m$ non-negative matrix $\Lambda$ are the term-topic loadings, the 
$m$-vector $\beta$ are regression coefficients that reveal the effect
of each topic on the response $Y$. 

To enhance interpretability of the model, we require that topic
loadings satisfy non-negativity constraints, which has been proposed
for matrix factorization with text and other forms of data in previous works, most notably
with extensions of the Nonnegative Matrix Factorization and Probabilistic Latent Semantic Analysis models \citep{NNMF1, NNMF2, NMF-PLSI, ConvexNMF}. 
The underlying intuition for why non-negativity is helpful with text is given in \cite{textNMF}. 
Documents and terms are grouped together by their underlying
topics and are also represented in the document-term
matrix as data points in the positive orthant. As a result, non-negativity constraints result in a factorization 
that is able to better match the geometry of the data by estimating
correlated vectors that identify each group of documents and terms. We build 
upon this literature and impose non-negativity to better capture the natural
geometry of the data. To understand the topic composition for a given document, one can
inspect the corresponding row of $X\Lambda$, where larger values
indicate greater topic importance to the document.  
  
Since the regression coefficients $\beta$ can take positive and negative values, the 
optimization problem most resembles the Semi-Nonnegative Matrix
Factorizations in \cite{ConvexNMF}, which was proposed for clustering
and visualization problems, and \cite{camsap2013-mankad-michailidis,
  mankad2014analysis}, who adapt the factorization for network
analysis. The exact form
and context of our model is, to our knowledge, novel, and manages to avoid the well-known issue of overfitting, which plagues other 
matrix factorization approaches in text analysis. Specifically, with classical techniques like 
Latent Semantic Analysis (see Section~\ref{sec:litreview} for detailed review; \cite{deerwester1990indexing}) or Probabilistic Latent Semantic Analysis \citep{pLSA}, 
one extracts topics by estimating a low-rank matrix factorization of the form $X \approx UDV^{T}$ subject to, respectively, the orthonormality constraints of 
Singular Value Decomposition or probability constraints. In both cases, 
the number of parameters grows linearly with the number of documents in the corpus. 
With the proposed factorization the number of parameters to estimate 
does not depend on corpus size, and grows with the size of the vocabulary and number of topics. 

We note that the factorization as posed above is not fully 
identifiable, as the columns of $\Lambda$ are subject to
permutations. The arbitrary ordering of topics is a feature present in all topic modeling techniques 
other than Latent Semantic Analysis. Moreover, note that 
$\Lambda D$ and $D^{-1}\beta$, where $D$ is a positive diagonal $m\times m$ matrix, is another solution with the same
objective value. 
We explored additional constraints on $\Lambda$ and/or $\beta$ to fix the scaling, but found that these approaches 
add complexity to the estimation without noticeably improving the quality of the final solution. Thus, we omit 
further discussion of these approaches here.

We also note that since the proposed method does not estimate a formal probability model for the topic structure, the document-term matrix $X$ can be preprocessed with term-frequency
inverse document frequency (TFIDF) weighting \citep{salton1983mcgill}
\begin{equation*}
(X)_{ij} = \mbox{TF}_{ij}\log(\frac{n}{\mbox{IDF}_{j}}), 
\end{equation*}
where $\mbox{TF}_{ij}$ denotes the term frequency (word count) of term $j$ in
document $i$, $\mbox{IDF}_{j}$ is the number of documents containing term
$j$, and $n$ is the total number of documents in the corpus. This
normalization has its theoretical basis in information theory and has
been shown to represent the data in a way that better discriminates
groups of documents and terms compared to simple word counts
\citep{robertson2004understanding}.

Finally, the proposed factorization can be used to generate predictions for any
new document by representing the document with the $p$-vector
$\tilde{x}$  so that the prediction is $\hat{y}=\tilde{x}\hat{\Lambda}\hat{\beta}$.

\subsection{Estimation} \label{sec:algo} 

The estimation approach we present alternates between optimizing with
respect to $\Lambda$ and $\beta$. The 
algorithm solves for $\Lambda$ using a projected gradient descent
method that has been effective at balancing cost per iteration and
convergence rate for similar problems posed in Nonnegative Matrix Factorization
\citep{lin2007projected}. 

Starting with $\beta$, when holding $\Lambda$ fixed, it is easy to verify that the remaining optimization 
problem is the usual regression problem leading to 
\begin{equation*}
\hat{\beta} = (\Lambda^{T}X^{T}X\Lambda)^{-1}\Lambda^{T}X^{T}Y.
\end{equation*}
Driven by our upcoming extension to the real data and results therein, we do not to regularize $\beta$, though it 
can be advantageous and easily done in other data contexts. 

Turning our attention to $\Lambda$, a standard gradient descent algorithm would start with an initial
guess $\Lambda^{(0)}$ and constants $\gamma_{i}$ and iterate:
\begin{enumerate} 
\item For $i=1,2,\ldots$
\item Set $\Lambda^{(i+1)} = \Lambda^{(i)} - \gamma_{i}\Delta_{\Lambda}$,
\end{enumerate}
where the gradient of the objective function with respect to $\Lambda$
is
\begin{equation} \label{eqn:gradientLambda}
\Delta_{\Lambda} = X^{T}X\Lambda\beta\beta^{T} - X^{T}Y\beta^{T}.
\end{equation}
Note that $X^{T}X$ and $X^{T}Y$ can be precomputed for faster
computing time. 

Due to the subtraction, the non-negativity of $\Lambda$ cannot be
guaranteed. Thus, the basic idea of projected gradient descent is to
project elements in $\Lambda$ to the feasible region using the
projection function, which for our problem is defined as $P(\gamma) =
max(0,\gamma)$. The basic algorithm is then
\begin{enumerate} 
\item For $i=0,1,2,\ldots$
\item Set $\Lambda^{(i+1)} = P(\Lambda^{(i)} - \gamma_{i}\Delta_{\Lambda})$.
\end{enumerate}

To guarantee a sufficient decrease at each iteration and convergence
to a stationary point, the ``Armijo rule'' developed in
\cite{bertsekas1976goldstein, bertsekas1999nonlinear} provides a
sufficient condition for a given $\gamma_{i}$ at each iteration
\begin{equation} \label{eqn:condition_orig}
||Y - X\Lambda^{(i+1)}\beta|| - ||Y - X\Lambda^{(i)}\beta|| \le \sigma \langle \Delta_{\Lambda^{(i)}},\Lambda^{(i+1)} - \Lambda^{(i)}\rangle,
\end{equation}
where $\sigma\in (0,1)$ and $\langle \cdot,\cdot \rangle$ is the sum
of element wise products of two matrices. Thus, for a given
$\gamma_{i}$, one calculates $\Lambda^{(i+1)}$ and checks whether
(\ref{eqn:condition_orig}) is satisfied.  If the condition is
satisfied, then the step size $\gamma_{i}$ is appropriate to guarantee
convergence to a stationary point. 

The final algorithm is given in pseudocode in 
Algorithm~\ref{algo:analysisprocedure:exact}. See the supplementary material \citep{supp_material} and Appendix~\ref{sec:app:algo1} for further discussion. 

\subsection{Extensions for Online Reviews Data: A Continuation Ratio Model with Embedded Topic Modeling} \label{sec:multinomial} 

In our data and generally with online review scores, $Y=\{1,2,3,4,5\}^{n}$, which are not well modeled with a normal distribution. 
To better fit our data, we embed the factorization within a type of ordinal regression, 
the continuation ratio model \citep[Ch.6]{fienberg2007analysis}, that incorporates time dynamics and multiple corpora (apps). 

We use the continuation ratio model instead of the more popular proportional odds model \citep{mccullagh1980regression} for primarily computational reasons, 
since the regression coefficients can be solved with standard logistic regression with the continuation ratio model. 
In practice, several researchers have observed that both forms of ordinal regression yield very similar results \citep{armstrong1989ordinal, archer20121, harrell2015regression}.
The basic idea is start with the following logit function $\mbox{logit}(Y = k) = \alpha_{k} + X\beta$, which we adapt to 
\begin{equation*}
\mbox{logit}(Y = k) = \alpha_{k} + X\Lambda\beta,
\end{equation*}
where $\mbox{logit}(Y=k) = \log\left(\frac{P(Y = k |Y \ge k, X)}{P(Y > k |Y \ge k, X)}\right)$. 
The corresponding likelihood is then the product of conditionally independent binomial terms for each level of $Y$. The log likelihood is given by 
\begin{eqnarray}
l(\Lambda, \beta|Y,X)  &=& \sum_{i=1}^{n}\sum_{k=1}^{K-1}(Y_{k})_{i}\log(p(k))  + (1 - \sum_{j=1}^{k}(Y_{j})_{i})\log(1 - p(k)) \nonumber \\ 
			    &=&  \sum_{i=1}^{n}\sum_{k=1}^{K-1}(Y_{k})_{i} \left(\alpha_{k} + (X)_{i}\Lambda\beta - \log (1 + e^{\alpha_{k} + (X)_{i}\Lambda\beta})\right) - \nonumber\\
			    &&   \sum_{i=1}^{n}\sum_{k=1}^{K-1}(1 - \sum_{j=1}^{k}(Y_{j})_{i}) \log(1 + e^{\alpha_{k} + (X)_{i}\Lambda\beta}), \nonumber
\end{eqnarray}
where $p(k)  =  P(Y_{i} = k | Y_{i} \ge k, (X)_{i}, \alpha_{k}, \beta) = \frac{e^{\alpha_{k} + (X)_{i}\Lambda\beta}}{1 + e^{\alpha_{k} + (X)_{i}\Lambda\beta}}$, $(X)_{i}$ refers to the $i$th row of $X$, and 
$Y_{k}$ are binary response vectors for categories $k=1,\ldots,K$ created from $Y$
\begin{equation*}
(Y_{k})_{j} =  \begin{cases}
	1 & \text{if } (Y)_{j} = k\\
	0 & \text{otherwise}
	\end{cases}
\end{equation*}
for $j=1,\ldots,n$ documents.

An important realization from the likelihood function is that it can be partitioned so that estimating the regression coefficients, holding $\Lambda$ fixed, can be done through standard binary logistic regression techniques. To our knowledge \cite{cox1988multinomial} and \cite{armstrong1989ordinal} were the first to show this for the standard continuation ratio model. The basic idea to apply logistic regression is to stack the recoded the response variables ($(Y_{k})_{j}$), including only observations that satisfy the condition $Y\ge k$ for $k=1,\ldots,K$, and duplicate corresponding rows to form the design matrix with dummy variables added to model the intercepts $\alpha_{k}$. In our context, the same trick can be applied when holding $\Lambda$ fixed. 

Recall our goal is to benchmark multiple apps over time, which calls for a dynamic model 
\begin{equation*}
\mbox{logit}(Y_{ta} = k) = \alpha_{tak} + X_{ta}\Lambda\beta_{ta},
\end{equation*}
where $a$ indexes the set of apps and $t$ denotes time.
Note that the number of documents changes with each app and time interval, but that the vocabulary is kept constant across them so that 
$X_{ta}$ is $n_{ta}\times p$, $Y_{tak}$ are $n_{ta}\times 1$ response vectors, and $\beta_{ta}$ are $m\times 1$ regression coefficients for each time interval, app category. 
Such a model is appropriate as long as the focal app or set of apps maintain the same core functionality, since then we could reasonably expect the discussion topics captured in $\Lambda$ to remain invariant. By visualizing $\beta_{ta}$ over time, as shown in Section~\ref{sec:results}, we can begin to understand the trend of consumer sentiment around topics in $\Lambda$ for different apps as well the effectiveness of development teams at responding to customer feedback. 

Another key assumption is that the regression coefficients $\beta_{ta}$ are independent of $k$, the rating level specified for each review. 
Arguably, this assumption is not germane to our online reviews data, since the occurrence and discussion of topics can have sentiment to them, and thus are related to the overall rating of the review. 
We also consider a saturated version of the model, where the regression coefficients vary with the level of the response variable
$\mbox{logit}(Y_{ta} = k) = \alpha_{tak} + X\Lambda\beta_{tak}$.
Likelihood ratio tests as well as out of sample prediction accuracy rates show that the constrained model is preferred, that is, assuming that $\beta_{tak} = \beta_{ta}$ for all $k$ leads to better statistical and predictive models (see Appendix~\ref{sec:app:podds} for more information). 

Estimation of the dynamic model follows a very similar alternating projected gradient descent algorithm as for the base factorization. When solving for $\Lambda$, holding $\alpha_{tak}$ and $\beta_{ta}$ fixed, we again utilize the projected gradient descent algorithm with appropriate updates for the gradient of $\Lambda$ and the Armijo rule \citep{bertsekas1976goldstein, bertsekas1999nonlinear} to guarantee convergence to a stationary point. Some further details are given in Appendix~\ref{app:fullalgo}. When holding $\Lambda$ fixed, one can estimate $\alpha_{tak}$ and $\beta_{ta}$ for each app-time by repeatedly utilizing the logistic regression solution from the static case for each app-time combination. To encourage smoothness in the regression coefficients, we utilize a rolling window so that $\alpha_{tak}$ and $\beta_{ta}$ are estimated using data from time points $t$ and $t-1$. Another approach yielding similar results would be to add a formal smoothness penalty to the log likelihood. A rigorous implementation of such an approach is outside the scope of this paper, but an interesting area of future work. 

Finally, when given a new document $x_{ta}$, one can predict the rating by selecting the response 
category with largest probability 
\begin{eqnarray}
P(Y_{ta}=1) &=& p(1)  \label{eqn:marginal1} \\
P(Y_{ta}=k) &=& p(k)\prod_{j=1}^{k-1} (1 - p(j)), k=2,\ldots, K-1 \\
P(Y_{ta}=K) &=& 1 - \sum_{k=1}^{K-1}P(Y_{ta}=k),   \label{eqn:marginal3}
\end{eqnarray}
where $p(k)  =  P(Y_{ta} = k |Y \ge k, x_{ta}, \Lambda, \alpha_{tak}, \beta_{ta}) = \frac{e^{\alpha_{tak} + x_{ta}\Lambda\beta_{ta}}}{1 + e^{\alpha_{tak} + x_{ta}\Lambda\beta_{ta}}}$.

\section{Relation with Topic Modeling Methods} \label{sec:litreview}
As shown in Table~\ref{table:litreview}, the historical roots of the proposed model go back to Latent Semantic Analysis (LSA), the most classical technique for
topic modeling, which is based on the Singular Value
Decomposition (SVD) of the document-term matrix $X \approx UDV^{T}$ \citep{deerwester1990indexing}. In many information retrieval tasks $X$ is projected onto the word-topic factors $XV^{T}$ for a low rank representation of the data. 
We of course are building on this idea with $X\Lambda$. With LSA, since $V$ can take elements of any sign, the interpretation of the resultant factors can be 
challenging in practice, which led to the development of the Probabilistic Latent Semantic Analysis. 

Probabilistic Latent Semantic Analysis (pLSA) developed in \cite{pLSA} 
is a formal probability model over the joint distribution of words and
documents. The idea is that each word in a document is a sample drawn
from a mixture of multinomial distributions that correspond to
different topics. pLSA can be written in the same algebraic 
form of SVD but imposes probability constraints, which greatly improved the interpretation of the resultant factors. 
In fact, \cite{NMF-PLSI} show an equivalency between the pLSA model
and the Non-Negative Matrix Factorization (NMF) of the document-term
matrix when one imposes sum
to one constraints in addition to the non-negativity for the NMF.  

While pLSA is widely seen as an improvement over LSA, there are two
major drawbacks.  First, the number of parameters to be estimated
grows linearly with the size of the corpus, which can lead to
overfitting.  Second, there is no systematic way to assign
probabilities to new documents after training the model.  
As discussed previously, both of these concerns are addressed in our model.

\begin{landscape}
\begin{table} 
\begin{tabular}{c|ccccc}
Method & \begin{tabular}{c}Decomposition\\Type\end{tabular} & Purpose & Supervised & \begin{tabular}{c}Incorporates\\Time\end{tabular} & \begin{tabular}{c}Multiple\\Corpora\end{tabular} \\\hline\hline
\begin{tabular}{c}Latent Semantic Analysis\\\citep{deerwester1990indexing}\end{tabular} & Orthonormal & Topic Modeling & No & No & No \\
\begin{tabular}{c}Probablistic Latent Semantic Analysis\\\citep{pLSA}\end{tabular} & Probabilistic & Topic Modeling & No & No & No \\
\begin{tabular}{c}Latent Dirichlet Allocation\\\citep{blei2003latent}\end{tabular} & Probabilistic & Topic Modeling & No & No & No \\
\begin{tabular}{c}Dynamic Latent Dirichlet Allocation\\\citep{blei2006dynamic}\end{tabular} & Probabilistic & Topic Modeling & No & Yes & No \\
\begin{tabular}{c}Supervised Latent Dirichlet Allocation\\\citep{mcauliffe2008supervised}\end{tabular} & Probabilistic & Topic Modeling \& Prediction & Yes & No & No \\
\begin{tabular}{c}Latent Aspect Rating Analysis\\\citep{wang2010latent}\end{tabular} & Probabilistic & Topic Modeling \&  Prediction  & Yes & No & No \\
\begin{tabular}{c}Multinomial Inverse Regression\\\citep{taddy2013multinomial}\end{tabular} & Logistic Regression & Sentiment Analysis & Yes & No & Yes\\
\begin{tabular}{c}Single Stage Matrix Factorization\\(Proposed Approach)\end{tabular} & Non-negative & Topic Modeling \& Prediction & Yes & Yes & Yes 
\end{tabular}
\caption{Summary and evolution of topic modeling methods.}
\label{table:litreview}
\end{table}
\end{landscape}

The latent
Dirichlet allocation (LDA) of \cite{blei2003latent} addresses these two 
issues with a hierarchical Bayesian generative model for how documents
are constructed. LDA has been shown to work very well in practice for
data exploration and unsupervised learning, and hence has been used
extensively in text mining applications \citep{blei2012probabilistic}. 
As mentioned previously, within the software quality and mobile app reviews literature, several papers (e.g., \cite{fu2013people, bavota2016mining}) use 
LDA as part of a multi-stage analysis that feeds into regression models and/or visualizations. 

We use the following LDA generating process in the next section to simulate documents in order to 
study how the proposed and competing methods perform in a controlled setting 
under various generating processes and signal-to-noise environments. 

The idea is that documents are constructed in a multi-stage procedure.
\begin{enumerate}
	\item Define $K$ topics, which are probability distributions over words and denoted as $\gamma_{1:K}$ .
	\item Randomly draw a distribution over topics for the entire corpus $\theta|\alpha \sim Dirichlet(\alpha)$.
	\item For each word in a document: 
	\begin{enumerate}
		\item Randomly sample a topic according to the distribution of topics created in Step 1, i.e., $z_{n}\sim Multinomial(\theta)$.
		\item Randomly sample a word according to the topic, i.e., $w_{n}|z_{n} \sim \gamma$. 
	\end{enumerate}	
\end{enumerate}
This generative process defines a joint probability
distribution, where the goal is to infer the conditional distribution
of the topic structure given the observed documents and word counts
\begin{equation*}
p(\gamma_{1:K},\theta_{1:D},z_{1:D}|w_{1:D}).
\end{equation*}
This task creates a key statistical challenge that has been addressed
with tools like Gibbs sampling \citep{Porteous:2008:FCG:1401890.1401960} 
or variational algorithms \citep{blei2006}. 

There have been several related extensions to LDA. For example, \cite{titov2008modeling} develop the Multi-grain Topic Model for modeling online reviews, which improves the coherence and interpretability of the topic-keywords by enforcing a hierarchical topic structure. The dynamic topic model \citep{blei2006dynamic} is another related extension that allows the topic loadings to change over time. These models do not consider document annotations or prediction, as in this work. 

The supervised latent Dirichlet allocation (sLDA) of
\cite{mcauliffe2008supervised} does consider document labels by adding a final stage to the LDA generative process, 
where a response variable is drawn on each document from the document's topic proportions.
\begin{enumerate}
	\item[4.] For each document, draw a response variable $Y|z_{1:N},\eta,\sigma^{2}\sim N(\eta^{T}\bar{z},\sigma^{2})$, where the prevalence of topics determine the outcome variable.
\end{enumerate}
sLDA has been utilized for recommender systems in the contexts of scientific articles \citep{wang2011collaborative} and physical products \citep{wu2015flame}, and 
extended to allow for additional covariates for the regression step \citep{agarwal2010flda}. 
We note that because these extensions are motivated by recommender systems, the focus is usually on adding latent variables 
that capture each user's affinity to different aspects of a product as he or she reviews items \citep{mcauley2013hidden}. Thus, conceptually the emphasis is on identifying preferences to products (or their attributes) at the user-level. 
Our work is motivated by a different problem that results in conceptual and modeling differences. Specifically, we are primarily interested in benchmarking from the product developer or designer's 
point of view, which requires understanding preferences at an aggregate (not user) level over time. Thus, one innovation we incorporate is to characterize the time evolution of how discussion 
on a common set of topic impacts the average customer's experience for multiple apps. This is an important extension, since this ultimately allows managers 
to go beyond a static summary of their app's performance to understand how the customer experience is evolving with different apps and versions. 
Additionally, because method does not estimate a formal probability distribution for the topic structure, we
can represent each document using the term-frequency inverse document frequency \citep{robertson2004understanding}, which has been shown to
be advantageous for various learning tasks. 
Our model also does not require tuning any parameters, whereas sLDA requires careful specification of hyperparameters. 
Numerous empirical studies show that the performance of LDA-based methods with online app reviews is sensitive to 
hyperparameter specification \citep{Lu2011, Panichella:2013:EUT:2486788.2486857, 6520844,  bavota2016mining}.

Another closely related literature stream is aspect modeling, where the main goal is to decompose a review into multidimensional aspects (topics) with ratings on each aspect \citep{titov2008joint}. Conceptually and at a high level, our work can be viewed as being representative of this stream, since in our model the $\Lambda$ and $\beta$ parameters encode, respectively, the ``aspects'' and their sentiment. The main difference between our work and the aspect modeling literature lies in the observable data structure and precise modeling goals. Most aspect modeling research assumes that ratings on each aspect are observable and have the goal of labeling each sentence within a review with an aspect and sentiment. Common modeling approaches are to extend LDA \citep{brody2010unsupervised, titov2008joint, lu2011multi, jo2011aspect} or pursue other similar latent variable models \citep{snyder2007multiple, brody2010unsupervised, mcauley2012learning}. For example, in our setting, the referenced aspect models would be appropriate if a reviewer provided separate numerical ratings for several dimensions, like functionality, user interface, reliability of the app, and so on.  However, this is rarely the case with app reviews, unlike reviews for restaurants on Yelp where such underlying aspects may be available. 

To our knowledge there is one work in aspect modeling that assumes an identical observable data structure. The Latent Aspect Rating Analysis model (LARA; \cite{wang2010latent}) aims to infer latent aspects and their sentiment scores from a review's text and its overall review rating. The paper follows a two-stage procedure, first using a seeded and iterative algorithm to identify aspects within each review, followed by a latent rating regression model. While LARA can be extended or modified to predict overall ratings, as in this work, the direct use-cases are distinct, namely annotation of sentences and inference of latent aspect ratings.

Finally we discuss the multinomial inverse regression of \cite{taddy2013multinomial}, which uses a logistic regression to extract sentiment information from document 
annotations and phrase counts that are modeled as draws from a multinomial distribution. 
The nuanced differences in context leads to different modeling decisions. 
Since sentiment analysis is the main objective in \cite{taddy2013multinomial}, where recovering dictionaries is critical, the multinomial inverse regression analysis is done at the phrase or term level. Our approach performs topic modeling (grouping of the terms) at the same time as regression.

\section{Simulation Study} \label{sec:simulation} 
We test the accuracy of the proposed model relative to competing methods  
under different settings. The first simulation 
establishes self-consistency of the proposed factorization, that is, responses are generated 
from the model implied by the factorization. 
The second simulation generates responses using the supervised latent Dirichlet allocation model 
of \cite{mcauliffe2008supervised}. For a fair comparison, we consider the canonical setting underlying (\ref{eqn:obj}) with a 
normally distributed response and without consideration of time or multiple apps. 

The methods we compare are as follows:
\begin{enumerate}
\item Latent semantic analysis of the document-term matrix with TFIDF
  weightings (denoted as LSA). Once the document term matrix has been
  decomposed with SVD, $X_{train} \approx UDV^{T}$, the singular
  vectors in $V$ are used as independent variables in a regression
  model $Y = X_{test}V\beta + \epsilon$;
\item Probabilistic latent semantic analysis (denoted as
  pLSA). Similarly, we estimate $Y = X_{test}V\beta + \epsilon$, where
  $V$ are the probabilistic word-topic
  loadings estimated from $X_{train}$;
\item Latent Dirichlet Allocation (LDA). Similarly, we estimate $Y = X_{test}V\beta + \epsilon$, where
  $V$ are the probabilistic word-topic loadings estimated from $X_{train}$. The Dirichlet parameters are chosen through five-fold cross validation;
\item Supervised LDA (denoted as sLDA). The Dirichlet parameters for the Document/Topic and Topic/Term distributions are chosen through five-fold cross validation and $\sigma^2$ is set to be the training sample variance;
\item $\ell_{1}$ penalized linear regression (Lasso; \cite{tibshirani1996regression, glmnet_jss}) of the response variable on the document term matrix. Ten-fold cross-validation on the training data is used to select the tuning parameter;
\item The proposed factorization of the document-term matrix (denoted as SSMF for Single-Stage Matrix Factorization).
\end{enumerate}

All analyses are performed using R \citep{R_Software}, with the ``tm''
\citep{tm_jss} and ``topicmodels'' \citep{R_topicmodels} libraries.
For sLDA, we use the collapsed Gibbs sampler implemented in the ``lda'' package
\citep{R_lda_package}. Code for the proposed Single-Stage Matrix
Factorization is provided in the supplementary material \citep{supp_material}.

\subsection{Self Consistency} 
Data are generated to study how the proposed model performs under its implied generating process, where 
$Y|X, \Lambda, \beta, \sigma^{2} \sim \text{Normal}(X\Lambda\beta,1)$. $X$ is the document term matrix,  
$(\Lambda)_{ij} \sim \text{Uniform}[0,1]$, and $(\beta)_{j} \sim \text{Normal}(0,1)$. 
Documents are simulated using the Latent Dirichlet Process \citep{blei2003latent} with both Dirichlet parameters for Document/Topic and Topic/Term distributions set equal to 0.8. 
The size of the vocabulary is set to $p=2000$ to roughly match our real dataset and others in the online review space \citep{buschken2016sentence, hospitality_spring}.

We vary the number of documents $n = \{100, 1000, 10000\}$ and the number of terms in each document $\mu=\{15, 250,2000\}$ to study how each model performs in different environments. The estimated number of topics is always equal to the true value and varied from $2$ to $20$. 
After training each model, we assess the accuracy of the predictions on the test set using the root mean squared error, which are shown in the top panel of Table~\ref{table:simulation}.

When the sample size is 1000 or lower, Lasso and the proposed model perform best. Lasso's performance is perhaps expected given that the generative model can be reparameterized as a linear regression  $X\Lambda\beta=X\gamma$, where $\gamma_{p \times 1}=\Lambda\beta$ is a vector of coefficients. It is notable that the proposed model performs well when the number of words in each document is small. 
This is important especially in the mobile apps context since the overwhelming majority of app reviews are written on mobile devices, leading to shorter and less formal writing styles  \citep{burtch2014happens}. In our real app reviews data, the average document length is under 20 words. When the number of documents is large, we see that all methods perform equally well, meaning that the advantages of supervision diminish in larger datasets.

% Requires LaTeX packages: dcolumn 
\begin{table}[!htbp]
\centering 
\begin{tabular}{c} 
Self-Consistency \\
\begin{tabular}{@{\extracolsep{0pt}} D{.}{.}{-3} D{.}{.}{-3} D{.}{.}{-3} D{.}{.}{-3} D{.}{.}{-3} D{.}{.}{-3} D{.}{.}{-3}D{.}{.}{-3} } \\[-1.8ex]\hline 
\\[-1.8ex]\hline 
\multicolumn{1}{c}{$\mu$} & \multicolumn{1}{c}{$n$} & \multicolumn{1}{c}{LSA} & \multicolumn{1}{c}{pLSA} & \multicolumn{1}{c}{LDA} & \multicolumn{1}{c}{sLDA}  & \multicolumn{1}{c}{Lasso}  & \multicolumn{1}{c}{SSMF} \\ 
\hline \\[-1.8ex] 
15 & 100 & 1.090  & 1.087  & 1.088  & 1.044  & 1.038  & 1.040  \\ 
& & (0.004) & (0.004) & (0.004) & (0.003) & (0.003) & (0.006) \\
15 & 1000 & 1.039  & 1.038  & 1.038  & 1.037  & 1.032  & 1.033  \\ 
& & (0.001) & (0.001) & (0.001) & (0.001) & (0.001) & (0.004) \\
15 & 10000 & 1.032 & 1.032  & 1.032  & 1.036  & 1.025  & 1.032  \\ 
& & (0.001) & (0.001) & (0.001) & (0.001) & (0.001) & (0.001) \\
250 & 100 & 1.056 & 1.058  & 1.057  & 1.323  & 1.018  & 1.009  \\ 
& & (0.003) & (0.003) & (0.003) & (0.012) & (0.003) & (0.003) \\
250 & 1000 & 1.007  & 1.007  & 1.007  & 1.013  & 1.003  & 1.003  \\ 
& & (0.001) & (0.001) & (0.001) & (0.001) & (0.001) & (0.001) \\
250 & 10000 & 1.002  & 1.002 & 1.002 & 1.004 & 1.002 & 1.002 \\ 
& & (0.001) & (0.001) & (0.001) & (0.001) & (0.001) & (0.001) \\
2000 & 100 & 1.050  & 1.049 & 1.049 & 1.710 & 1.011 & 0.999 \\ 
& & (0.003) & (0.003) & (0.003) & (0.028) & (0.003) & (0.003) \\
2000 & 1000 & 1.004  & 1.004 & 1.004 & 1.027 & 1.001 & 1.001\\ 
& & (0.001) & (0.001) & (0.001) & (0.002) & (0.001) & (0.001)\\
2000 & 10000 & 0.998  & 0.998 & 0.998 & 1.002 & 0.999 & 0.998 \\ 
& & (0.002) & (0.002) & (0.002) & (0.002) & (0.002) & (0.002) \\
\hline \\[-1.8ex] 
\end{tabular}  \\
sLDA Generating Process \\
\begin{tabular}{@{\extracolsep{0pt}} D{.}{.}{-3} D{.}{.}{-3} D{.}{.}{-3} D{.}{.}{-3} D{.}{.}{-3} D{.}{.}{-3} D{.}{.}{-3}D{.}{.}{-3} } \\[-1.8ex]\hline 
\hline \\[-1.8ex] 
\multicolumn{1}{c}{$\mu$} & \multicolumn{1}{c}{$n$} & \multicolumn{1}{c}{LSA} & \multicolumn{1}{c}{pLSA} & \multicolumn{1}{c}{LDA} & \multicolumn{1}{c}{sLDA}  & \multicolumn{1}{c}{Lasso} & \multicolumn{1}{c}{SSMF} \\ 
\hline \\[-1.8ex] 
15 & 100 & 1.110 & 1.109 & 1.112  & 1.077  & 1.073  & 1.106  \\ 
 & & (0.005) & (0.005) & (0.005) & (0.005) & (0.004) & (0.013)\\
15 & 1000 & 1.057  & 1.057  & 1.057  & 1.056  & 1.052 & 1.053 \\ 
 & & (0.002)  & (0.002) & (0.002) & (0.002)  & (0.002) & (0.008) \\
15 & 10000 & 1.051  & 1.051  & 1.051  & 1.053  & 1.051  & 1.052  \\ 
& & (0.002) & (0.002) & (0.002) & (0.002) & (0.002)  & (0.002) \\
250 & 100 & 1.102  & 1.102  & 1.102  & 1.293  & 1.077  & 1.077  \\ 
& & (0.005) & (0.005) & (0.005) & (0.012) & (0.005) & (0.006) \\
250 & 1000 & 1.068  & 1.068  & 1.068  & 1.070  & 1.069  & 1.068  \\ 
& & (0.004) & (0.004) & (0.004) & (0.004) & (0.004) & (0.004) \\
250 & 10000 & 1.054 & 1.053  & 1.054  & 1.059  & 1.053  & 1.050 \\ 
& & (0.002) & (0.002) & (0.002) & (0.002) & (0.002) & (0.002) \\
2000 & 100 & 1.100  & 1.100  & 1.099  & 1.730  & 1.206  & 1.060  \\ 
 & & (0.005) & (0.005) & (0.005) & (0.035) & (0.025) & (0.005) \\
2000 & 1000 & 1.072 & 1.072  & 1.072  & 1.087  & 1.095  & 1.070  \\ 
& & (0.004) & (0.004) & (0.004) & (0.004)  & (0.011) & (0.004) \\
2000 & 10000 & 1.001  & 1.001 & 1.001 & 1.002 & 0.999 & 1.000 \\ 
& & (0.002) & (0.002) & (0.002) & (0.002) & (0.002) & (0.002) \\
\hline \\[-1.8ex] 
\end{tabular} 
\end{tabular} 
  \caption{Root Mean Squared Error averaged over all ranks from the simulation study with standard errors in parentheses.} 
  \label{table:simulation} 
\end{table} 

\subsection{Supervised Latent Dirichlet Allocation} 
Data are generated under the generating process assumed by sLDA \citep{mcauliffe2008supervised}, where 
$Y|Z,\beta,\sigma^{2} \sim \text{Normal}(\beta^{T}Z,\sigma^{2})$.  $Z$ is the Document/Topic probability distribution. 
All other settings are identical to the previous simulation study. Table~\ref{table:simulation} shows that sLDA and Lasso perform 
best with a $n=100$ and $\mu=15$, with the proposed method coming in third. In other settings every method tends to perform similarly. 
The robust performance of SSMF in both simulations with documents of varying length indicates that the proposed factorization should be useful for our app review data as well as with other corpora. 

\section{iTunes and Google Play App Reviews} \label{sec:results} 

We now demonstrate the 
method's real-life viability and applicability by using the mobile apps marketplace data 
from the apps provided by Expedia, Kayak, and TripAdvisor that we described earlier. 
We begin by discussing the preprocessing and model selection 
steps, followed by a detailed discussion of the findings. 

To ensure accurate word
counts when forming the document term matrix, we follow the standard preprocessing steps \citep{boyd2014care} of
transforming all text into lowercase and removing punctuation, stopwords (e.g., ``a'', ``and'', ``the''), and any terms composed of less
than three characters. In addition to counting the frequency of single words, we also count bigrams, which are all two word phrases that appear in the corpus. 
For example, the sentence ``this is a wonderful app'' is tokenized into single words ``this'', ``is'', ``a'', ``wonderful'', ``app'' as well as two-word phrases ``this is'', ``is a'', ``a wonderful'', and ``wonderful app''.  After counting all unigrams and bigrams, we remove terms that have occurred in less than 20 reviews and apply TFIDF weighting. 
The resulting total vocabulary size is 2583 for reviews from iTunes and 1389 for reviews from Google Play.  

\begin{table}
\begin{tabular}{l|rrcc}
Platform-App & \begin{tabular}{c}Number\\Reviews\end{tabular} & \begin{tabular}{c}Average\\Review Length\\(characters)\end{tabular} & \begin{tabular}{c}\% Reviews\\with ``!''\end{tabular} & Yule's K \\\hline 
\\[-1.8ex]\hline 
iTunes-Expedia & 2772 & 98.715 & 28.968 & 62.325 \\
iTunes-Kayak & 13120 & 68.948 & 31.623 & 59.289 \\ 
iTunes-TripAdvisor & 19519 & 107.949 & 32.235 & 71.308 \\\hline
Google Play-Expedia & 6999 & 95.246 & 15.416 & 69.915 \\
Google Play-Kayak & 21059 & 58.023 & 15.267 & 49.637 \\ 
Google Play-TripAdvisor & 41347 & 65.660 & 14.069 & 53.895
\end{tabular}
\caption{Summary statistics for the online reviews data. Yule's K is a measure of vocabulary richness, where higher numbers indicate a more diverse vocabulary \citep{holmes1985analysis}.}
\label{table:realdata_summarystats}
\end{table}

Table~\ref{table:realdata_summarystats} shows an overview of the review data, where we see that despite being a younger platform, Google Play has more reviews for every app. The customer writing style also seems to vary by platform. 
iTunes reviews tend to be longer, potentially more emotional due to greater number of exclamation points, and have a higher lexical diversity. We analyze each platform separately due to these differences in addition to the fact that the hardware (mobile phones and tablets) that run the mobile apps vary across platforms, as do the underlying development enviroments that used to develop code for the apps. We also define time in terms of year-quarters in our analysis to avoid sparsity issues early in an app's lifecycle and also to roughly match the approximate rate at which upgrades and new functionalities are released for the apps in our sample. The last observed quarter for each platform is withheld as the test set. 

Cross-validation applied to the training sample selects five topics for iTunes and four topics for the Google Play platform according to misclassification error rate (MER). Table~\ref{table:top10_Final} shows the top ten keywords from 
our final models using. Each of the topics were manually labeled with headings after inspecting the keywords and reviews that loaded most heavily onto each topic. For instance, Table~\ref{table:reviews_GPlay} shows reviews that correspond to the largest values in columns (topics) of $X\Lambda$ for the Google Play data. Due to space constraints, the top reviews for all topics and the iTunes data are omitted.

%The number of topics selected corresponds roughly to the number of dimensions underpinning
%software quality theory \citep{kitchenham1996software, jung2004measuring[[, and we can see some of these fundamental dimensions in topics.]]}. The usual theoretical dimensions are %functionality, reliability, usability, efficiency, maintainability, and portability.
%In the mobile app context, we expect reviewers to comment on functionality, usability, and efficiency since these concepts are user-facing.

\begin{table}
\begin{tabular}{c}
iTunes \\ %\hline\\[-1.8ex]\hline 
\begin{tabular}{p{3.1in}|p{3.1in}}
\hspace{0.8in} Topic 1 Keywords & \hspace{0.8in} Topic 2 Keywords \\
\hspace{0.475in} Usability (Online Reviews) & \hspace{0.4in}Functionality (Reservations) \\\hline
useful, helpful, good, cool, awsome, nice, great app, availability, great, wish app & 
indispensable, reviewers, advisor always, since last, app also, establishment, helpfull, properties, helping, reviews pictures \\
\hspace{0.8in} Topic 3 Keywords & \hspace{0.8in} Topic 4 Keywords \\
\hspace{0.95in}Overall Quality & \hspace{1.2in}Versioning \\\hline
great, awesome, love, easy, app, best, use, amazing, great app, perfect, easy use &  
fill, forced, changing, worthless, latest version, returned, happened, old version, back old, bring back  \\
\hspace{0.8in} Topic 5 Keywords \\
\hspace{0.35in}Functionality (Software Bugs) \\\hline
emails, crashes, almost every, dont want, one star, category, glitch, apply, internet connection, customer service
\end{tabular}\\\\
GooglePlay \\ %\hline\\[-1.8ex]\hline 
\begin{tabular}{p{3.1in}|p{3.1in}}
\hspace{0.8in} Topic 1 Keywords & \hspace{0.8in} Topic 2 Keywords \\
\hspace{0.4in} Functionality (Reservations) & \hspace{0.6in} Usability (UI \& Design) \\\hline
brilliant, comment, paid, wasnt, seriously, scroll, coupon, hotel flight, apparently, main  & 
helpful, half, average, expensive, enter, agent, availability, advertised, liked, order \\
\hspace{0.8in} Topic 3 Keywords & \hspace{0.8in} Topic 4 Keywords \\
\hspace{0.35in}Usability (Composing Reviews) &  \hspace{0.55in} Installation \& Versioning \\\hline
write reviews, find way, asking, poor, line, app im, searched, app book, either, downloading & 
bloatware, stupid, uninstalled, uninstall, useless, crap, remove, month, return, expensive, message
\end{tabular}
\end{tabular}
\caption{The top ten topic keywords from estimating five topics on iTunes and four topics on Google Play.}
\label{table:top10_Final}
\end{table}

\begin{landscape}
\begin{table}
\begin{tabular}{p{0.95\columnwidth}}
\textbf{Reviews that load most heavily onto ``Usability (Composing Reviews)''}\\\hline\\[-1.8ex]\hline 
\begin{tabular}{p{0.95\columnwidth}}
``i downloaded app on my new phone to write reviews  i was able to find what i wanted to write reviews and entering the review was easy i havent searched for lists yet''\\
``just cant submit have to paste cant type into input line for submit name''\\
``on several occasions reviews that i have submitted have either failed to submit successfully and give an error message or have mysteriously disappeared after apparently being submitted successfully   another variant problem is that draught that are saved can also disappear  it is extremely frustrating to spend perhaps  minutes writing a review on a mobile device only to find that the time has been wasted  in terms of displaying tripadvisor information on the move it is reasonable''\\
``but in entering a comment i went up to add something and it wouldnt allow me to scroll back down to complete my comment  so i had to either completely redo the review or just enter and hope for the best  since the livelihood of establishments depend upon these comments this should not happen  please fix it  this happened on a galaxy tablet''
\end{tabular}\\
\textbf{Reviews that load most heavily onto ``Installation \& Versioning''}\\\hline\\[-1.8ex]\hline 
\begin{tabular}{p{0.95\columnwidth}}
``new mobile tablet app is frustrating unable to book hotel reservation with more than one traveler unable to book multiple rooms unable to make changes in reservation unable to cancel reservation through mobile app unable to use online support telephone support slow and useless credit for first time mobile use unavailable after telephone support call hang up dial the hotel direct and dump this app''\\
``i cant uninstall this app on my s i dont have a need for it please let me remove it im so sick of bloatware apple has the right idea when it comes to controlling what goes on there phonessmh sprint and samsung you make it hard to be a loyal customer when there are apps on my phone that i dont need or use that cant be uninstalled''\\
``samsung should stop adding crap bloatware and leave us course what we want installed on our phone  why we cant remove such app bring the phone with the app installed but leave us remove it i never use such app and i dont need it''\\
``cant uninstall cant remove''
\end{tabular}
\end{tabular}
\caption{Top reviews associated with different topics for the Google Play platform.}
\label{table:reviews_GPlay}
\end{table}
\end{landscape}

Assessing the quality of topic keywords can be challenging, since interpretability is a difficult characteristic to quantify. 
\cite{mimno2011} provide one solution in a measure called topic coherence, where the general idea is to 
gauge the interpretability of each topic based on co-occurrences of its keywords. 
The average keyword coherence is defined as 
\begin{equation*}
\mbox{Coherence} = \frac{2}{Kp(p-1)}\sum_{k=1}^{K}\sum_{u=2}^{p}\sum_{v=1}^{u-1}\log\left(\frac{D(w_{u}^{k},w_{v}^{k})+0.01}{D(w_{v}^{k})}\right),
\end{equation*}
where $(w_{1}^{k},\ldots,w_{p}^{k})$ is the list of $p$ top words in topic $k$, $K$ is the number of topics,  $D(w)$ is the number of reviews containing the word $w$, and $D(w,w')$ is the number of reviews containing both $w$ and $w'$. The constant $0.01$ is added to avoid taking the log of zero when two keywords do not co-occur over all documents. Coherence is bounded above by zero; model results with larger  coherence scores have been shown to be more interpretable by human judges \citep{mimno2011}. 
To measure redundancy of the recovered topics, we report Uniqueness, which is defined as the average proportion of keywords in each topic that do not appear as keywords for other topics (similar to  ``inter-topic similarity'' in \cite{arora2013practical}). Larger values indicate more useful results. The top and middle panels of Table~\ref{table:realdata_errorrates} shows that the proposed method is generating interpretable and useful results. In contrast to competing methods that tend to score well on either Coherence or Uniqueness, SSMF is competitive on both dimensions.

A third way to validate our results is to compare out of sample forecasts. We generate predictions on the test set by using the estimated 
document-topic matrix $\hat{\Lambda}$ and regression coefficients from the most recent quarter $\hat{\beta}_{T-1,a}$. 
We again benchmark the performance against LSA, pLSA, LDA, sLDA, and Lasso. The two-stage procedures utilize the continuation ratio model in the second stage and Lasso refers to the $\ell_{1}$ penalized continuation ratio model of \cite{archer20121}. We also include a standard continuation ratio model with all unigrams and bigrams as covariates and no penalty. The bottom panel of Table~\ref{table:realdata_errorrates} shows that Lasso and the proposed method produce the most accurate predictions. These results are consistent with the simulation study that showed these two methods performing well among the tested methodologies when the sample size is in the thousands, which approximately matches the number of reviews received each quarter collectively for the three apps.

\begin{landscape}
\begin{table}
\centering 
\begin{tabular}{c}
Keyword Coherence \\ 
\begin{tabular}{c|ccccccc}
Platform & LSA & pLSA & LDA & sLDA & Lasso & \begin{tabular}{c}Ordinal\\Regression\end{tabular} & SSMF \\\hline\\[-1.8ex]\hline 
iTunes & -0.566 & -0.750 & -0.564 & -0.688 & NA &  NA & -0.743\\
           & (0.355\%) & (32.979\%) & max &  (27.035\%) & & & (26.017\%)\\ 
Google & -0.862 & -1.113 & -0.826 & -1.057 & NA &  NA & -0.926\\
Play        & (4.358\%) & (34.746\%) & max &  (27.966\%) & & & (12.107\%)\\ 
\end{tabular}\\
Uniqueness \\ 
\begin{tabular}{c|ccccccc}
Platform & LSA & pLSA & LDA & sLDA & Lasso & \begin{tabular}{c}Ordinal\\Regression\end{tabular} & SSMF \\\hline\\[-1.8ex]\hline 
iTunes & 0.096 & 0.726 & 0.132 & 0.344 & NA &  NA & 0.592 \\
           & (86.777\%) & max & (81.818\%) &  (52.617\%) & & & (18.457\%)\\ 
Google & 0.188 & 0.843 & 0.170 & 0.475 & NA  & NA & 0.580\\
Play        & (77.699\%) & max & (79.834\%) & (43.654\%) &  &  & (31.198\%)\\
\end{tabular}\\
Misclassification Error Rates \\ 
\begin{tabular}{c|ccccccc}
Platform & LSA & pLSA & LDA & sLDA & Lasso & \begin{tabular}{c}Ordinal\\Regression\end{tabular} & SSMF \\\hline\\[-1.8ex]\hline 
iTunes & 0.319 & 0.313 & 0.320 & 0.712 & 0.294 & 0.319 & 0.299 \\
       & (8.503\%) & (6.463\%) & (8.844\%) & (142.180\%) & min  & (8.503\%)  & (1.701\%)\\
Google & 0.373 & 0.376 & 0.376 & 0.639 & 0.334 & 0.377 & 0.327 \\
  Play     & (14.067\%) & (14.984\%) & (14.984\%) & (95.413\%) & (2.140\%)  & (15.291\%) & min\\
\end{tabular}
\end{tabular}
\caption{Average topic coherence and uniqueness based on the top 100 topic keywords, and out of sample misclassification error rate when predicting online review ratings in the first quarter of 2014 for iTunes and third quarter of 2014 for Google Play. The reported percentage is the relative 
    percentage of difference to the best result. Note that sLDA was run assuming a normally distributed response as this is the only working option in the public code. All other methods were combined with a continuation ratio model.}
\label{table:realdata_errorrates}
\end{table}
\end{landscape}

\begin{figure}
\begin{tabular}{c}
iTunes\\
\includegraphics[width=1\columnwidth, trim=0cm 0.2cm 0cm 0cm, clip=true]{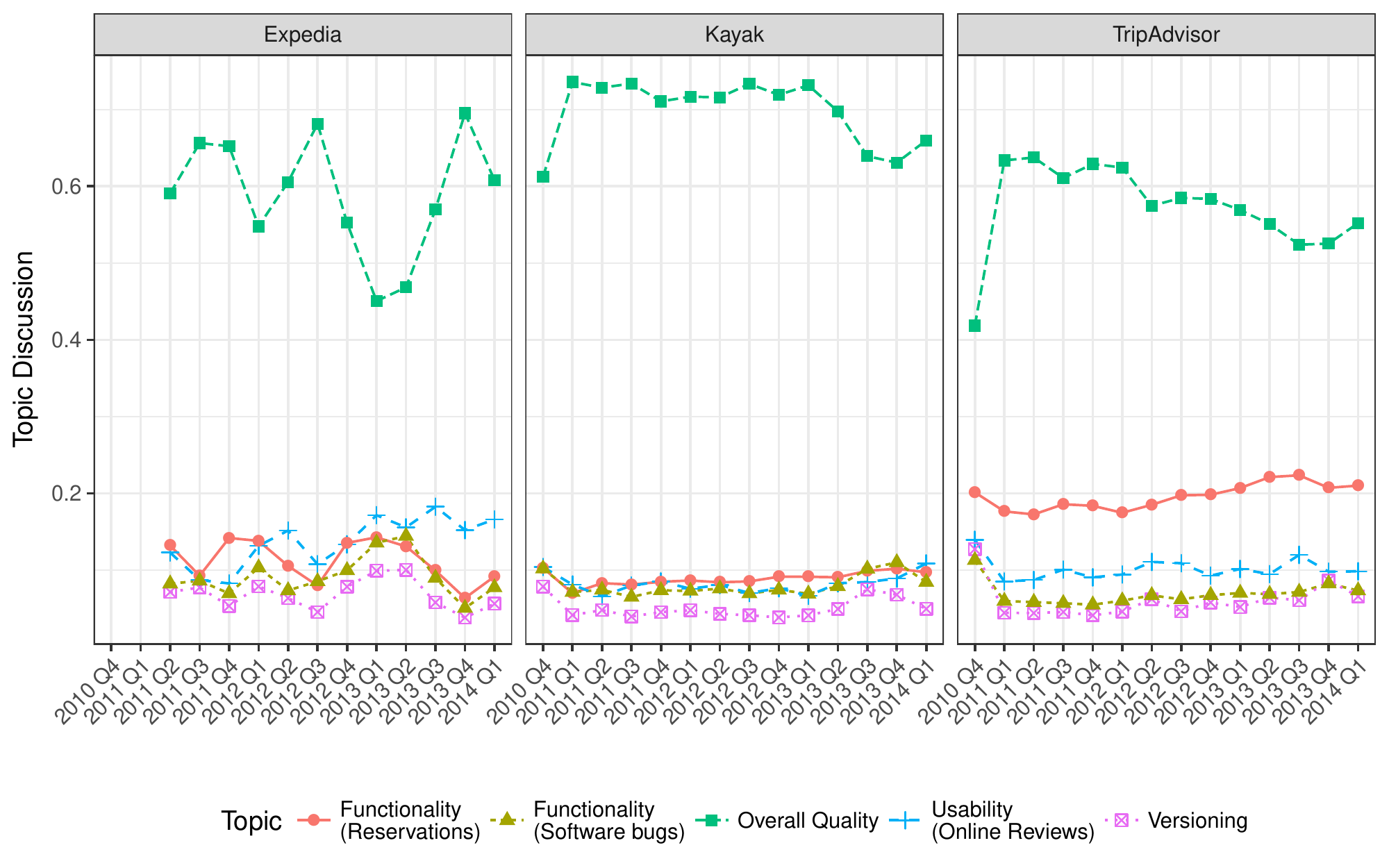}\\
Google Play \\
\includegraphics[width=1\columnwidth, trim=0cm 0.2cm 0cm 0cm, clip=true]{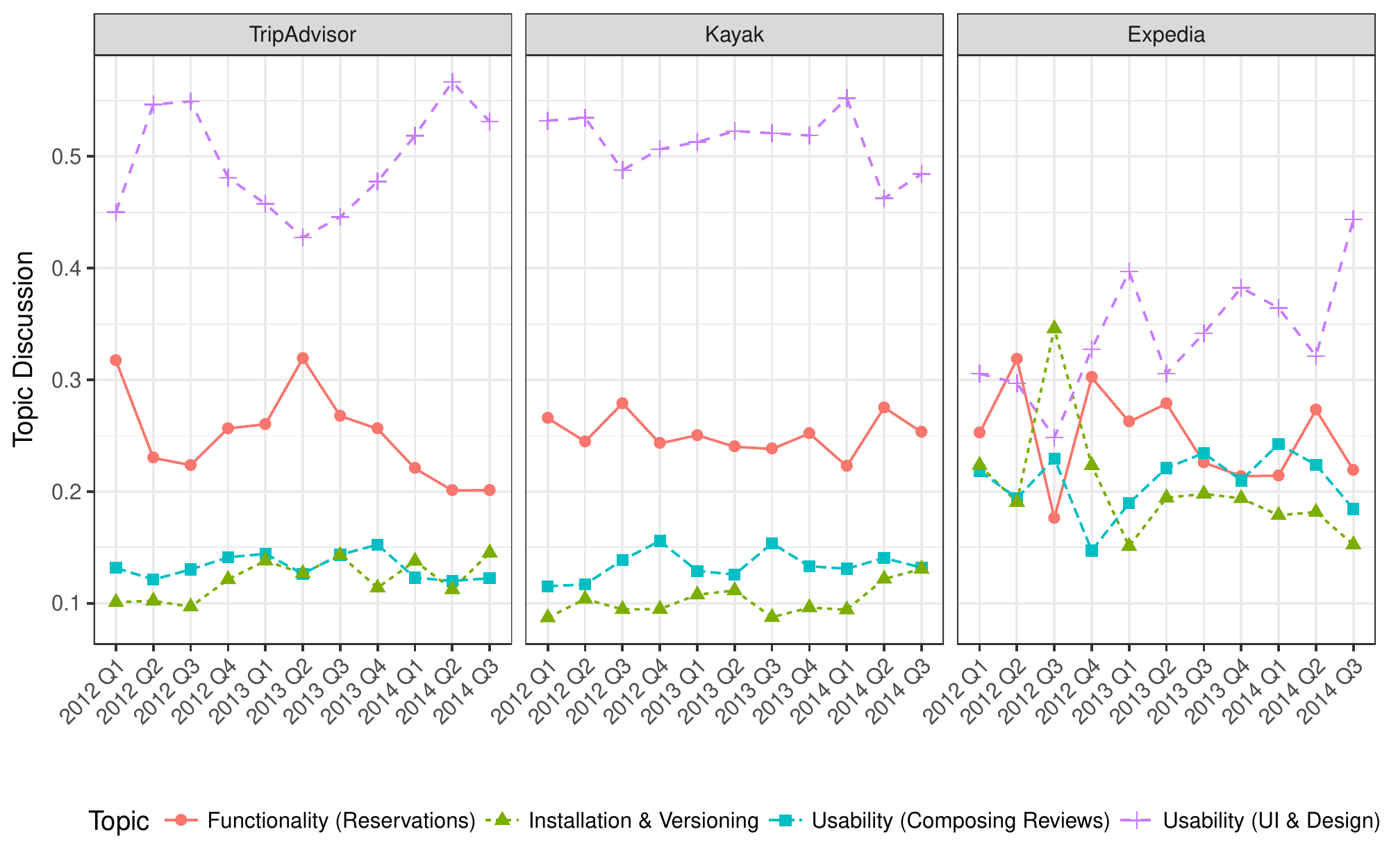}
\end{tabular}
\caption{Prevalence of topics in reviews over time.}
\label{fig:loadings_overtime}
\end{figure}

\begin{figure}
\begin{tabular}{c}
iTunes\\
\includegraphics[width=0.85\columnwidth, trim=0cm 2.1cm 0cm 0cm, clip=true]{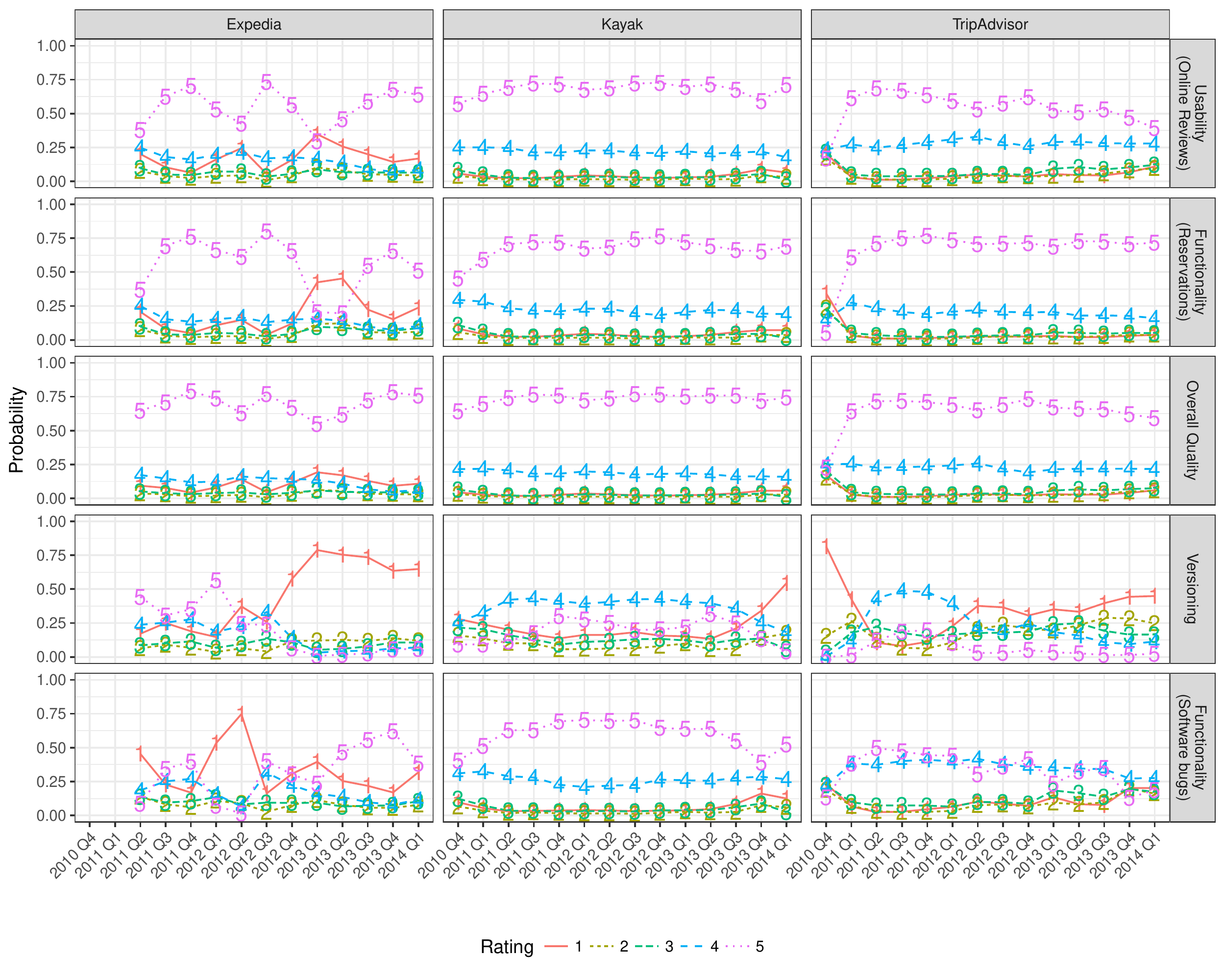}\\
Google Play \\
\includegraphics[width=0.85\columnwidth]{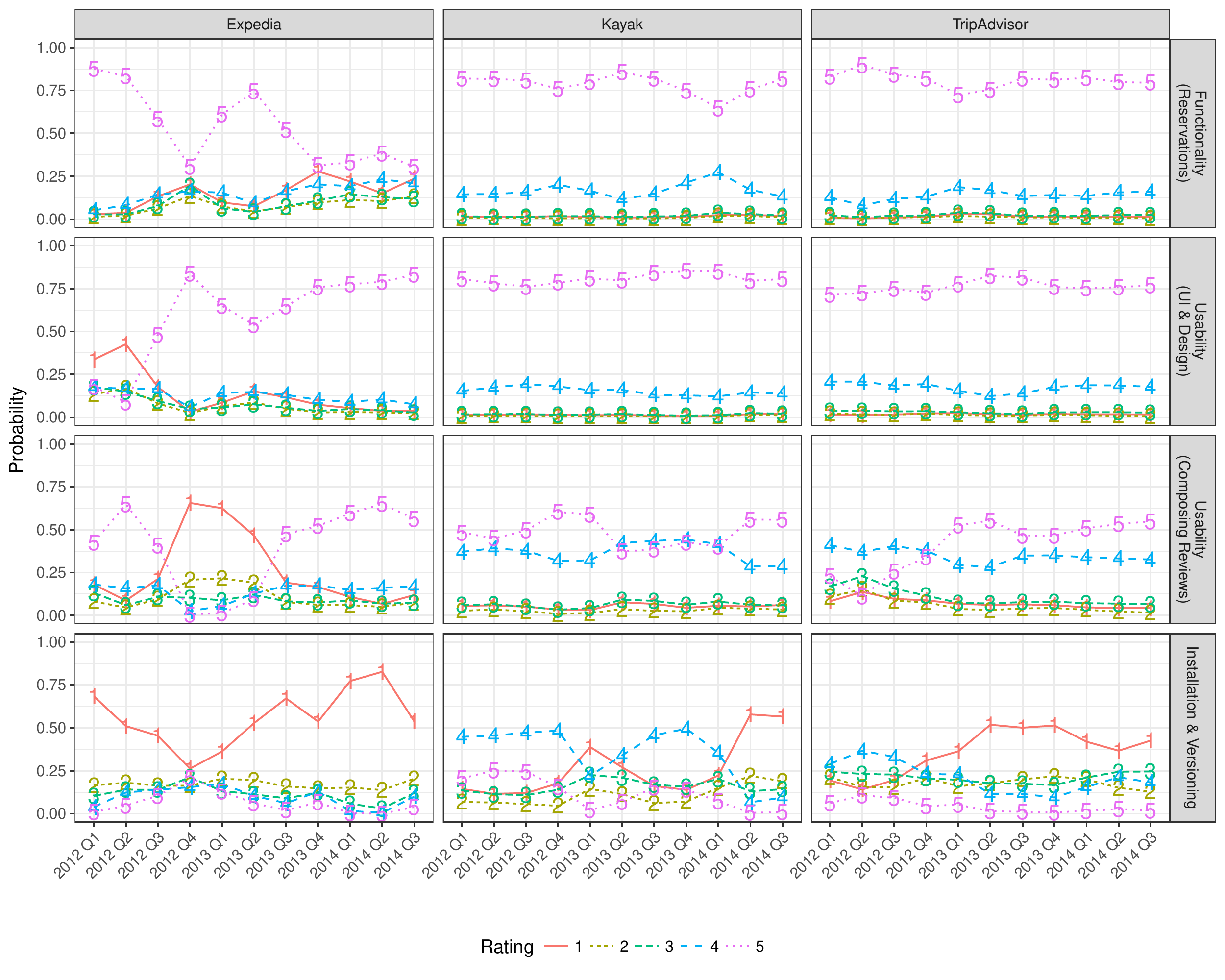}
\end{tabular}
\caption{Probability of ratings by topic appearance.}
\label{fig:probabilities_overtime}
\end{figure}

Having chosen and validated the proposed models, we turn to synthesizing our findings from the mobile apps data and the the estimation of $\Lambda$ and $\beta$, which are summarized in Figures~\ref{fig:loadings_overtime} and \ref{fig:probabilities_overtime}, respectively. 
Figure~\ref{fig:loadings_overtime} shows the amount of discussion in each quarter on each topic, assessed by taking column sums of $X_{ta}\Lambda$. Figure~\ref{fig:probabilities_overtime} displays the regression coefficients transformed into probabilities, which is necessary to avoid interpretation difficulties that arise with viewing the coefficients directly. Specifically, $P(Y_{ta} = k)$ is calculated by considering a hypothetical document that loads onto a single topic, where $X_{ta}\Lambda = e_{m}$ and $e_{m}$ is a vector with 1 in the $m$-th position and zero elsewhere. The required marginal probabilities can be readily computed using (\ref{eqn:marginal1})-(\ref{eqn:marginal3}). 
 
These two figures, combined with the ratings evolutions in Figure~\ref{fig:tsOverview}, show several interesting patterns that help characterize the evolution of each app over time while also identifying areas of improvement for the respective app development teams. 
We see a small dip in the overall ratings for the Kayak app on iTunes in the third and fourth quarter of 2013. This decrease coincided with discussion around two issues: software bugs (crashes, API errors, etc.) and versioning, which are then associated with higher chances of the app being rated lower on the 5-point scale by users.  Even though the volume of discussion was fairly stable, the topics became increasingly toxic as users were rating the app more harshly along these dimensions, thereby dragging down the overall rating. Similarly, we can see the odds of receiving 1-star reviews strongly increasing with the occurrence of these topics within reviews, coinciding with a negative episode in the overall ratings for the Expedia app on iTunes between the third quarter of 2012 and the third quarter of 2013. 
In fact, we can see from Figure~\ref{fig:probabilities_overtime} that Expedia has persistent problems with versioning and software bugs on both platforms that are on-going at the end of the data. On Google Play, Expedia is generally rated lower than its competitors, and we see that, in addition to versioning, the company had difficulty especially in 2012 with general user interface issues around the launch of the app, followed by difficulties around composing and posting online reviews by its users. In contrast, TripAdvisor has consistently been rated highly on both platforms since the apps were introduced to the public. Interestingly, on both platforms we see installation and versioning as significant sources of discontent from users, though the amount of discussion on these topics has been  low. Nonetheless, it is relevant to note that TripAdvisor forces automatic updates of its apps on both platforms and is even embedded in the operating system as a default program on certain versions of Android mobile phones, which is at the heart of the negative feedback from users. This raises an interesting tradeoff for TripAdvisor's mobile strategy - between the options of increasing its user base by being embedded within the Android system versus the cost of alienating some users who may be annoyed at having to uninstall the app manually. 

\section{Conclusion} \label{sec:conclusion} 

Consider a mobile app developer who has introduced an app on the Google Play app store and has received, over a period of time, several thousand reviews from users. 
Ideally, the developer would like to extract some information from these reviews that will help inform where the main problems are with the developed app, as well as where the app stands with respect to competitor apps on dimensions that relate to user experience or service quality. 
Furthermore, over time, the developer would like to understand time trends relating to dimensions of feedback from online reviews, and how these are associated with the received app rating. 
In this paper, we present an ordinal regression framework with embedded topic modeling  
to recover topics from online reviews that are predictive of the star rating in addition to being
useful for understanding the underlying textual themes. 
Moreover, this model performs particularly well in the specific context of mobile apps, 
where reviews tend to be short, change over time with app versions, but have common elements in terms of what users tend to discuss in these reviews. 

We demonstrated how the model can be applied for benchmarking by analyzing mobile app reviews for Expedia, Kayak, and TripAdvisor. 
Specifically, by investigating the trend in overall ratings in combination with the estimated ratings probability by topic, 
we identified potential reasons behind poor user satisfaction that resulted in negative movements in the overall star rating for an app.
For instance, we observe that the increased odds of receiving 1-star reviews during the final quarter of our dataset is associated with negative feedback related to the topic described by versioning issues. On deeper examination, we conclude that all three companies on the two platforms should be particularly cognizant of the pros and cons of the versioning strategy they espouse. 
While forcing users into app updates may help improve the user experience for some users by fixing software bugs or introducing new and important features to the updated app, 
there is the potential cost of alienating a different and potentially overlapping set of users when such automatic updates are too frequent or add low quality features. Such a potential tradeoff can be deduced by app developers through the use of the SSMF approach we describe. 
In a related manner, and as discussed above for TripAdvisor, a similar potential downside from a development perspective exists with respect to the strategy of preinstalling the app on mobile phones. While this strategy helps some users, it can cause dissatisfaction to others who are faced with having to delete the app manually. In yet another instance during the final quarter of data, Expedia's iTunes-based users report the presence of critical software bugs while Android-based users complain about the reservation functionality. Based on our methodology, it would be possible for Expedia to corroborate these initial insights through traditional software testing and redirect their app development team's efforts more effectively towards tackling these sources of discontent among its users. 

It is interesting to note that our proposed model and Lasso generally performed the best, and on par with each other, among the tested methodologies in terms of forecasting accuracy on both the real reviews data as well as on simulated data. These results are consistent with \cite{o2015analysis} who showed that NMF style factorizations 
may lead to better solutions compared to LDA-based approaches, especially with 
niche or non-mainstream corpora,  such as reviews for mobile apps on mobile devices, which tend to be short and informal. 
Another factor determining the efficacy of the proposed model, relative to other models, is sample size  (number of documents). 
In the simulation, Lasso and the proposed models were preferred when the sample size was in the thousands or smaller. 
At larger sample sizes in each time point, our simulation indicated that two-stage procedures with standard topic modeling 
in the first stage perform equally well. 

While we consider three clearly competitive apps within the same industry here, an important and particularly insightful extension of our methodology could be to recover market structure for the entire app market using online app reviews. Market structure is an important factor in firm-level decision making pertaining to product development, pricing, and marketing strategies. Yet, in general with mobile apps, the appropriate set of benchmark or competitive apps is unclear, especially from the consumers's perspective. For instance, if an app streams video even without it being a core feature, the average consumer might benchmark this functionality internally against 
Netflix or the YouTube app, popular apps that specialize in video playback. Thus, identifying which other apps are seen by the consumer as competitors or substitutes could be derived from the set of online reviews associated with each of these apps, thereby enhancing the value that companies gain from a better understanding of online reviews. Tackling this problem would likely require analyzing data from a much 
broader set of mobile apps, potentially the entire marketplace, which raises several methodological issues from preprocessing the data \citep{fu2013people} to summarizing network structure 
and trends over time. 
As such, a growing number of firms have begun developing dashboards that display summaries of online customer reviews to managers \citep{hospitality_spring}. Our methodology is promising for such summaries that require benchmarking, understanding market dynamics, and prediction accuracy.
%Thus, our methodology offers promise beyond simply informing the focal app developer about the insights latent in mobile reviews. Our work described here also offers the potential to enhance the process of benchmarking and establishing competitive categories within the mobile app ecosystem, thereby allowing from higher prediction accuracy, greater efficacy in allocating development effort, and a deeper understanding of market dynamics in the fast-moving world of mobile apps. 

\section*{Acknowledgements}
This material is based upon work supported by the National Science Foundation under Grant No. 1633158 (Mankad).  

The authors would also like to thank the anonymous referees, the Associate Editor, and Editor for constructive comments and suggestions that resulted in a much improved paper.

\begin{supplement} \label{suppA}%[id-suppA] 
%\sname{Supplement A} 
\stitle{Raw Data and R Code}
\slink[doi]{COMPLETED BY THE TYPESETTER}
\sdatatype{.zip}
\sdescription{The zip file contains the raw online reviews data for the three apps on both platforms in addition to implementations in R of the proposed matrix factorization.}
\end{supplement}

\bibliographystyle{imsart-nameyear}
%\bibliography{biblio}

\appendix 

\section{Algorithm for the Single Stage Matrix Factorization with Normal Responses} \label{sec:app:algo1} 

The final algorithm for the SSMF is given in 
Algorithm~\ref{algo:analysisprocedure:exact}.  

Updating $\Lambda$ and specifically searching for an
appropriate $\gamma_{i}$ when updating $\Lambda$ is the most
time-consuming task. 
The major computation when searching for a good step size is $\langle \Delta_{\Lambda^{(i)}},\Lambda^{(i+1)} - \Lambda^{(i)}\rangle$,
where $\langle \cdot,\cdot \rangle$ is the sum of element wise products of two matrices. 

Breaking down this specific calculation, we focus on the gradient which is defined in (\ref{eqn:gradientLambda}).
$X^{T}X$, $\beta\beta^{T}$, and $X^{T}Y\beta^{T}$ can all be precomputed before entering into the step size search. In fact, 
$X^{T}X$ and $X^{T}Y$ can be computed before beginning Algorithm~\ref{algo:analysisprocedure:exact}. Due to these precomputations, the cost of searching for the step size is 
\begin{equation*}
\underbrace{\mathcal{O}(p^2 n) + \mathcal{O}(pn) + \mathcal{O}(pm) + \mathcal{O}(m^2)}_{\text{Precomputed $X^{T}X$, $X^{T}Y$, $(X^{T}Y)\beta^{T}$, and $\beta\beta^{T}$}} + \#\text{sub-iterations} \times (\underbrace{\mathcal{O}(p^2 m)}_{(X^{T}X)\Lambda} + \underbrace{\mathcal{O}(pm^2)}_{\Lambda(\beta\beta^{T})}).
\end{equation*}

Adding in the cost of the element-wise sum and estimating $\beta$ with standard procedures ($\mathcal{O}((n+m)m^2)$), the overall cost of the algorithm is 
\begin{flalign}
&\mathcal{O}(p^2 n) + \mathcal{O}(pn) +& \nonumber\\
&\#\text{iterations} \times \left(\mathcal{O}((n+m)m^2) + \mathcal{O}(pm) + \mathcal{O}(m^2) + \#\text{sub-iterations} \times \left(\mathcal{O}(p^2 m) + \mathcal{O}(pm^2)\right)\right).& \nonumber
\end{flalign}

\begin{figure}
\centering \includegraphics[width=0.7\columnwidth]{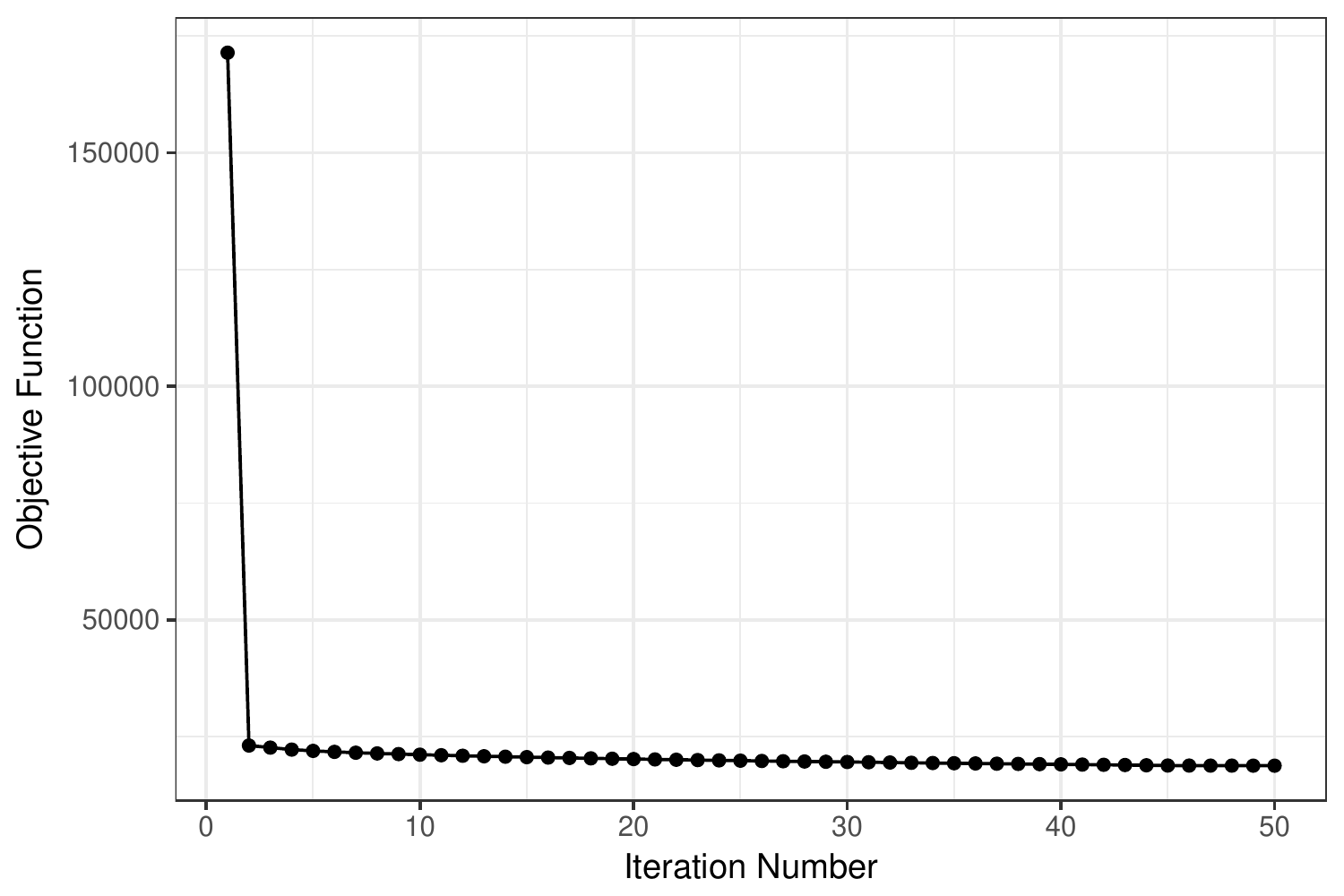}
\caption{One instance of the objective function at each iteration of the SSMF estimation. 
The alternating projected gradient descent algorithm monotonically improves the estimates with respect to the objective function.}
\label{fig:objfn}
\end{figure}

As long as the number of sub-iterations is small, the algorithm is efficient for the given data, especially since the vocabulary size is not extremely large. 
To this end, we utilize the heuristic of
using $\alpha_{i-1}$ as an initial guess for $\gamma_{i}$, and set $\sigma=0.01$ and $\gamma = 0.9$.  
Figure~\ref{fig:objfn} shows the algorithm results in estimates that monotonically improve at each iteration and converge fairly quickly. In our experiments, the relative difference between objective values converged to within $10^{-4}$ typically within 15 iterations.

\begin{algorithm}
\begin{algorithmic}[1]
\STATE Set $i = 0$ 
\STATE Initialize $(\beta)_{j}^{(i)} \sim N(0,1)$ for all $j$ 
\STATE Initialize $\gamma_{i}=1, \gamma=0.9$ 
\WHILE {$\delta \ge \epsilon$ \AND $i \le \text{max iterations}$} 
	\STATE $\gamma_{i+1} = \gamma_{i}$
	\IF {$\gamma_{i+1}$ satisfies (\ref{eqn:condition_orig})}
		\REPEAT \STATE{$\gamma_{i+1} = \frac{\gamma_{i+1}}{\gamma}$}
		\UNTIL{$\gamma_{i+1}$ does not satisfies (\ref{eqn:condition_orig})}
	\ELSE 
		\REPEAT \STATE{$\gamma_{i+1} = \gamma_{i+1}\gamma$}
		\UNTIL{$\gamma_{i+1}$ satisfies (\ref{eqn:condition_orig})}
	\ENDIF
	\STATE Set $\Lambda^{(i+1)} = P\left(\Lambda^{(i)} - \gamma_{i+1}(X^{T}X\Lambda\beta\beta^{T} - X^{T}Y\beta^{T})\right)$
	\STATE Set $\tilde{X} = X\Lambda^{(i+1)}$. 
	\STATE Set for $\beta^{(i+1)}=(\tilde{X}^{T}\tilde{X})^{-1}\tilde{X}^{T}Y$.
	\STATE Set $\delta =  \frac{||Y - X\Lambda^{(i+1)}\beta^{(i+1)}||_{2}^{2} - ||Y - X\Lambda^{(i)}\beta^{(i)}||_{2}^{2}}{||Y - X\Lambda^{(i)}\beta^{(i)}||_{2}^{2}}$ 
	\STATE Set $i =  i + 1$ 
\ENDWHILE
\end{algorithmic}
\caption{The Alternating Least Squares Algorithm with projected
  gradient descent for normally distributed $Y$, where the superscript $(i)$ denotes the iteration
  number.}
\label{algo:analysisprocedure:exact}
\end{algorithm}

\section{Comparing the Constrained and Saturated Continuation Ratio Models} \label{sec:app:podds} 

In this section we evaluate whether the constrained or saturated model is preferred. The results presented here use the real data from the app marketplaces and the final proposed model that includes regression coefficients varying over time and app.  In this framework, the constrained model specifies that $\beta_{tak} = \beta_{ta}$ for all $k$. 

We compare the nested models using likelihood ratio tests. Define the likelihood ratio statistic
\begin{equation*} 
G = 2\left(l(\text{Saturated Model}) - l(\text{Constrained Model})\right),
\end{equation*}
following a Chi-Squared distribution with $df_2 - df_1$ degrees of freedom, where 
\begin{eqnarray*}
df_{1} &=& \# \text{Topics} * \# \text{Apps} * \#\text{Time points} \\
df_2 &=& \# Topics * \# Apps *  \#\text{Time points} * (\# \text{ Rating categories}-1).
\end{eqnarray*}
On the iTunes data the likelihood ratio statistic $G= 12.575$ has a p-value close to $1.000$ and on Google Play $G=423.811$ has a p-value of $0.161$. 
Failing to reject the null hypothesis on both platforms indicates that the constrained model fits as well as the saturated version. Thus, we prefer the constrained version of the model.

This decision is confirmed by the out of sample misclassification error rates on our online reviews data in Table~\ref{table:podds} . 
The constrained model performs favorably, especially on the Google Play data, indicating that the more complex, saturated model likely overfits the data. 

\begin{table}
\centering\begin{tabular}{l|cc}
Platform & Saturated SSMF & SSMF \\\hline\\[-1.8ex]\hline 
iTunes & 0.300 & 0.299\\
Google Play & 0.337 & 0.327 \\  
\end{tabular}
\caption{Out of Sample Misclassification Error Rates of the proposed model with regression coefficients that vary over time and app.  The Saturated SSMF allows the regression coefficients to additionally vary for each category versus fixed over ratings categories.}
\label{table:podds}
\end{table}

\section{Estimation of the Dynamic Factorization Embedded Continuation Ratio Model} \label{app:fullalgo}

The log-likelihood function for the proposed model is 
\begin{eqnarray}
l(\Lambda, \beta_{ta}|Y_{tak},X_{ta})  &=& \sum_{t=1}^{T}\sum_{a=1}^{A}\sum_{i=1}^{n_{ta}}\sum_{k=1}^{K-1}(Y_{tak})_{i}\log(p(k))  + (1 - \sum_{j=1}^{k}(Y_{taj})_{i})\log(1 - p(k)) \nonumber \\ 
			    &=&  \sum_{t=1}^{T}\sum_{a=1}^{A}\sum_{i=1}^{n_{ta}}\sum_{k=1}^{K-1}(Y_{tak})_{i} \left(\alpha_{tak} + (X_{ta})_{i}\Lambda\beta_{ta} - \log (1 + e^{\alpha_{tak} + (X_{ta})_{i}\Lambda\beta_{ta}})\right) - \nonumber\\
			    && (1 - \sum_{j=1}^{k}(Y_{taj})_{i}) \log(1 + e^{\alpha_{tak} + (X_{ta})_{i}\Lambda\beta_{ta}}). \nonumber
\end{eqnarray}

When solving for $\Lambda$, holding all other parameters fixed, we again utilize the projected gradient descent algorithm with appropriate updates for the gradient of $\Lambda$ and the Armijo rule shown below. 

The gradient of the log-likelihood with respect to $\Lambda$ is
\begin{eqnarray*}
\Delta_{\Lambda} = \frac{\partial l}{\partial \Lambda} &=& \sum_{t=1}^{T}\sum_{a=1}^{A}\sum_{i=1}^{n_{ta}}\sum_{k=1}^{K-1} (Y_{tak})_{i} \frac{1}{1 + e^{\alpha_{tak} + (X_{ta})_{i}\Lambda\beta_{ta}}}(X_{ta})_{i} ^{T}\beta_{ta}^{T} +\\
&& (1 - \sum_{j=1}^{k}(Y_{taj})_{i})\frac{-e^{\alpha_{tak} + (X_{ta})_{i}\Lambda\beta_{ta}}}{1 + e^{\alpha_{tak} + (X_{ta})_{i}\Lambda\beta_{ta}}}(X_{ta})_{i} ^{T}\beta_{ta}^{T}.
\end{eqnarray*}

To guarantee a sufficient decrease at each iteration and convergence
to a stationary point, the Armijo rule is used to select appropriate $\gamma_{i}$ at each iteration 
\begin{equation*} %\label{eqn:condition_new}
l(\Lambda^{(i+1)},\beta_{ta}|Y_{ta},X_{ta}) - l(\Lambda^{(i)},\beta_{ta}|Y_{ta},X_{ta}) \le \sigma \langle \Delta_{\Lambda^{(i)}},\Lambda^{(i+1)} - \Lambda^{(i)}\rangle,
\end{equation*}
where $\sigma\in (0,1)$ and $\langle \cdot,\cdot \rangle$ is the sum
of element wise products of two matrices. 

\end{document}